\documentclass[11pt]{article}
\usepackage{amssymb}
\usepackage{amsmath}
\usepackage[all]{xy}
\input{epsf}
\usepackage{epsfig}
\numberwithin{equation}{section}

\def\be{\begin{equation}}
\def\ee{\end{equation}}
\def\ba{\begin{align}}
\def\ea{\end{align}}
\def\beq{\begin{eqnarray}}
\def\eeq{\end{eqnarray}}

\def\p{\partial}

 \input{epsf}

 \usepackage{epsfig}

\begin{document}

\title{\Large{\bf 
The influence of D-branes' backreaction upon gravitational interactions between open strings
}} 
\author{Raphael Benichou 
}
\date{}
\maketitle
\begin{center}
  Theoretische Natuurkunde, Vrije Universiteit Brussel and \\
The International Solvay Institutes,\\
 Pleinlaan 2, B-1050 Brussels, Belgium \\
 \textsl{raphael.benichou@vub.ac.be}
\end{center}

\begin{abstract}
We argue that gravitational interactions between open strings ending on D3-branes are largely shaped by the D3-branes' backreaction.
To this end we consider classical open strings coupled to general relativity in Poincar\'e $AdS_5$ backgrounds.
We compute the linear gravitational backreaction of a static string extending up to the Poincar\'e horizon, and deduce the potential energy between two such strings.
If spacetime is non-compact, we find that the gravitational potential energy between parallel open strings is independent of the string's inertial masses and goes like $1/r$ at large distance $r$.
If the space transverse to the D3-branes is suitably compactified, a collective mode of the graviton propagates usual four-dimensional gravity.
In that case the backreaction of the D3-branes induces a correction to the Newtonian potential energy that violates the equivalence principle.
The observed enhancement of the gravitational attraction at arbitrary large distances is specific to string theory; there is no similar effect for point-particles.

\end{abstract}

\newpage

\tableofcontents

\newpage


\section{Introduction}

D-branes \cite{Polchinski:1995mt} have played an important role in the recent developments of string theory.
D-branes can be described using either an open or a closed string language.
In the open string sector D-branes are boundary conditions for the worldsheet conformal field theory.
In the closed string sector D-branes are sources for the closed string fields, and consequently they are associated with solutions to the equations of motion of supergravity.
In a regime in which the bulk closed strings decouple from the open strings living on the D-branes, these two points of view give equivalent descriptions of the same objects.
This observation lead to the formulation of the AdS/CFT correspondence \cite{Maldacena:1997re}\cite{Aharony:1999ti}.
On the other hand when the open and closed strings sectors are coupled together we have to take into account both aspects of the D-branes physics.

D-branes are also useful objects for model-building in string theory.
Generically at low energy the theory living on D-branes is a gauge theory. 
One can build D-branes configurations such that the resulting low-energy effective theory resembles the Standard Model of particles physics, or a supersymmetric extension thereof. 
This gauge theory is coupled to gravity which is mediated by closed strings living in the bulk.
These models also offer an interesting perspective to address cosmological questions such as inflation.

Let us consider a configuration of D-branes arranged on a manifold $\mathcal{M}$.
In some regime it may be legitimate to treat the D-branes as probes. 
Then the gravitational sector of the low-energy effective theory is (a supersymmetric extension of) general relativity living on $\mathcal{M}$, and coupled to the gauge theory.
When the D-branes' backreaction is taken into account this picture has to be modified.
Indeed gravitons being elementary excitations of the bulk geometry, they are sensitive to the D-branes' backreaction.
The goal of this paper is to investigate how the gravitational sector of the low-energy effective theory is affected by the D-branes' backreaction.

Let us introduce a motivation for this work.
It is a question of importance to identify observable features of string theory that would not be reproduced by a theory of point particles.
From the point of view of a distant observer, when a string approaches a gravitational horizon it becomes tensionless and its proper size grows.
Thus in the neighborhood of horizons strings and point-particles should behave quite differently.
D-branes are heavy objects that create a horizon.
Open strings attached to D-branes stretch up to the D-branes horizon.
Consequently their proper length is typically infinite.
So we may expect that interactions mediated by bulk degrees of freedom have different manifestations for open strings and for point particles, even at macroscopic distances.
The results we will derive indicate that it is indeed the case.

In this paper we focus on open strings attached to D3-branes spanning four-dimensional Minkowski spacetime.
The space transverse to the D3-branes may be compact or non-compact; we will deal with both cases.
D3-branes create a coordinate horizon.
The classical gravity solution for D3-branes does not contain any curvature singularity and can be continued beyond the horizon \cite{Gibbons:1994vm}. 
Here we adopt the view that spacetime finishes at the horizon, where the D3-branes sit. Any closed string that crosses the horizon is turned into a set of open strings attached to the D3-branes (see e.g. \cite{Callan:1996dv} for an early development of this picture in the context of black holes thermodynamics).
Open strings attached to the D3-branes extend up to the horizon,
and their proper length is infinite (see section \ref{NGString} for explicit computations).

The main result of this paper is the computation of the classical gravitational potential energy between two open strings attached to a stack of D3-branes.
We use an effective field theory to describe gravity in the bulk. 
The open strings are described by the Nambu-Goto action and couple to gravity via their stress-energy tensor.
As long as the open strings are not heavily excited their worldsheets are localized close to the D3-branes. 
We assume that the near-horizon geometry of the stack of D3-branes is the product of the Poincar\'e patch of $AdS_5$ with a compact manifold:
\be\label{ds2spacetime} ds^2 =  \frac{\rho^2}{R^2} \eta_{\mu \nu} dx^\mu dx^\nu + \frac{R^2}{\rho^2} d\rho^2 +  ds^2_{compact}\ee
where $R$ is the $AdS_5$ radius.
The D3-branes live at $\rho=0$, where the metric develops a horizon.
The prototypal example is the case of D3-branes located on a smooth ten-dimensional manifold.
The compact space is then a five-sphere of radius $R$, which takes the value: 
\be\label{radius} R^2 = \sqrt{4\pi g_s N} \alpha' \ee
where $g_s$ is the string coupling, $\alpha'$ is the string length squared and $N$ is the number of D3-branes.
We assume that the $AdS_5$ radius $R$ is large with respect to the string scale $\sqrt{\alpha'}$.
In this regime we can trust the field theory description of gravity.
It is convenient to work with the coordinate $z = \frac{R^2}{\rho} $, so that the metric of the Poincar\'e patch of $AdS_5$ is conformally flat\footnote{We adopt the convention that Greek indices $\mu,\nu,...$ denote four-dimensional Minkowski coordinates. Latin indices $a,b...$ are associated with five dimensional coordinates $(z,x^\mu)$.}:
\be\label{Poincare} ds^2 =  R^2 \left(\frac{\eta_{\mu \nu} dx^\mu dx^\nu + dz^2}{z^2}  \right) \ee
The horizon and the stack of D3-branes are now located at $z = \infty $.

In order to simplify the computations we will first discard the compact factor in the metric \eqref{ds2spacetime} and work in five dimensions. 
In section 2 we describe linearized gravity in Poincar\'e $AdS_5$. 
In section 3 we compute the linear gravitational backreaction of an open string that stretches up to the Poincar\'e horizon. 
In section 4 we evaluate the gravitational potential energy between two such strings. 
We show that this potential energy is independent of the open strings' inertial masses and goes at large distance like $1/r$, where $r$ is the distance between the two strings as measured by a Minkowski observer.
We propose the following interpretation for this result:
the dominant contribution to the gravitational interaction comes from the parts of the open strings located very close to the horizon, in a region where the geodesic distance between the strings is small and the effect of curvature is negligible. Accordingly the open strings interact like infinite strings in five-dimensional flat space.

In section \ref{generalizationDim} we discuss the influence of the compact factor in the metric \eqref{ds2spacetime}.
In section \ref{generalization} we generalize our results to compact spacetimes.
Eventually section \ref{discussion} contains a summary and a discussion of the results, as well as directions for future research.


\section{Linearized gravity in Poincar\'e $AdS_5$}\label{linGrav}


\subsection{Linearized Einstein's equations}

The Poincar\'e $AdS_5$ metric \eqref{Poincare} satisfies Einstein's equations in the vacuum with appropriate negative cosmological constant. 
In this section we study the linearized Einstein's equations for a small perturbation of the metric \eqref{Poincare}.
We follow the analysis of \cite{Giddings:2000mu} (see also \cite{Garriga:1999yh}).
We work with the action:
\be S = \int d^{5}x \sqrt{-g} \left( \frac{1}{G_N^{(5)}} \mathcal{R} - \Lambda + \mathcal{L}_{matter} \right) \ee
where $G_N^{(5)}$ is the five-dimensional Newton's constant, $\mathcal{R}$ is the Ricci scalar, $\Lambda$ is the cosmological constant and $\mathcal{L}_{matter}$ is the Lagrangian describing the matter sources.

Let us introduce a small perturbation in the metric \eqref{Poincare}.
We can always choose a coordinate system such that the perturbed metric takes the form :
\be\label{PertPoincare} ds^2 =  \frac{dz^2 + \left( \eta_{\mu \nu} + h_{\mu \nu}(z,x^\rho) \right) dx^\mu dx^\nu}{z^2} \ee
We treat the metric perturbation $h_{\mu \nu}$ as a small fluctuation.
Linearized Einstein's equations read :
\begin{align}\label{linEqs} & -\mathcal{R}^{(4)} - \frac{3}{R^2} z \p_z h = \frac{z^2}{R^2} G_N^{(5)} T_{zz} \cr
& \p_z \p^\nu (h_{\mu \nu} - h \eta_{\mu \nu}) = G_N^{(5)} T_{z \mu} \cr
& \mathcal{G}^{(4)}_{\mu \nu} + \frac{3z}{2R^2}  \p_z(h_{\mu \nu} - \eta_{\mu \nu} h) - \frac{z^2}{2R^2}\p^2_z(h_{\mu \nu} - \eta_{\mu \nu} h) = \frac{z^2}{R^2}  G_N^{(5)} T_{\mu \nu}
\end{align}
where the indices $\mu,\nu$ are lowered and raised with the $4$-dimensional Minkowski metric $\eta_{\mu \nu}$,  $h = h_{\mu \nu} \eta^{\mu \nu}$, and $\mathcal{R}^{(4)}$ and ${\mathcal{G}^{(4)}}^\mu_\nu$ are respectively the linearized part of the Ricci scalar and Einstein's tensor built with the $4$-dimensional metric $\frac{R^2}{z^2} \left( \eta_{\mu \nu} + h_{\mu \nu} \right)$.

We will now show that we can essentially reduce the linearized Einstein's equations \eqref{linEqs} to a scalar wave equation in $AdS_5$. 
First let us take the trace of the last line in \eqref{linEqs} and combine it with the first line to eliminate the Ricci scalar. We obtain:
\be\label{linEqH} \frac{z}{R} \p_z \left(\frac{R}{z} \p_z h\right) = \frac{G_N^{(5)}}{3} \left( \eta^{\mu \nu} T_{\mu \nu} - 2T_{zz} \right) \ee
This equation is easily integrated to get $\p_z h$, which can be set to zero outside the sources with a reparametrization of the form:
\be z \to z - \frac{z}{R} \alpha(x^\nu) \qquad ; \qquad x^\mu \to x^\mu + \frac{z^2}{2 R}  \p^\mu \alpha(x^\nu) \ee
Such a reparametrization preserves the form of the metric \eqref{PertPoincare}. 
Next let us define $\bar h_{\mu \nu}$ as:
\be\label{defBarH} \bar h_{\mu \nu} = h_{\mu \nu} - \frac{1}{2} \eta_{\mu \nu} h \ee
The second line in \eqref{linEqs} is then:
\be\label{linEqH2} \p_z \p^\nu \bar h_{\mu \nu} = \frac{1}{2}\p_z \p_\mu h + G_N^{(5)} T_{z \mu} \ee
This equation is in turn straightforwardly integrated to obtain $\p^\nu \bar h_{\mu \nu}$, which can also be set to zero outside the sources with a reparametrization of the form:
\be z \to z  \qquad ; \qquad x^\mu \to x^\mu - \beta^\mu (x^\nu) \ee
Eventually the last line in \eqref{linEqs} reads :
\be \label{linEqHmunu} \Box \bar h_{\mu \nu}  = \frac{z^2}{R^2} \left( - \eta_{\mu \nu} \p^\lambda \p^\sigma \bar h_{\lambda \sigma} + \p^\lambda \p_\mu \bar h_{\nu \lambda} + \p^\lambda \p_\nu \bar h_{\mu \lambda} \right) 
 + \frac{\eta_{\mu \nu}}{2} \frac{z^{5}}{R^{5}} \p_z \left( \frac{R^3}{z^3} \p_z h \right) 
 - \frac{z^2}{R^2} G_N^{(5)} T_{\mu \nu} \ee
where $\Box$ is the anti-de Sitter Laplacian.
%
%
The right-hand side is known explicitly in terms of the matter stress-energy tensor thanks to equations \eqref{linEqH} and \eqref{linEqH2}, together with the gauge choice that both $\p_z h$ and $\p^\nu \bar h_{\mu \nu}$ vanish away from the sources.


\subsection{Green functions}\label{green}

In this section we construct a Green function that allows for the integration of equation \eqref{linEqHmunu} 
(see for instance \cite{Giddings:2000mu}\cite{Garriga:1999yh}\cite{Kiritsis:2003mc}\cite{Rubakov:2001kp} for similar discussions).
Let us introduce the functions $\phi_{\mu \nu}$ defined as:
\be\label{phi(h)} \phi_{\mu \nu}= \left(\frac{z}{R}\right)^{-\frac{3}{2}} \bar h_{\mu \nu} \ee
The virtue of this field redefinition is that it transforms the $AdS_5$ wave equation into a Schr\"odinger-like equation:
\be \Box \bar h_{\mu \nu} = \left(\frac{z}{R}\right)^{\frac{7}{2}} L\ \phi_{\mu \nu} \ee
where $L$ is the following linear differential operator:
\be\label{opL} L = \p^2_z - \frac{15}{4 z^2} + \eta^{\mu \nu} \p_\mu \p_\nu \ee
So equation \eqref{linEqHmunu} can be rewritten as:
\be\label{linEqPhimunu}  L\ \phi_{\mu \nu} 
 = \left(\frac{z}{R}\right)^{-\frac{3}{2}}\left[ \left( - \eta_{\mu \nu} \p^\lambda \p^\sigma \bar h_{\lambda \sigma} + \p^\lambda \p_\mu \bar h_{\nu \lambda} + \p^\lambda \p_\nu \bar h_{\mu \lambda} \right) 
 +  \frac{\eta_{\mu \nu}}{2} \frac{z^{3}}{R^{3}} \p_z \left( \frac{R^3}{z^3} \p_z h \right) 
 -  G_N^{(5)} T_{\mu \nu} \right] \ee
Let us introduce the Green function $G(x^a,{x'}^a)$ for the operator $L$:
\be\label{defG} L\ G(x^a,{x'}^a) = \delta^{(5)}(x^a-{x'}^a) \ee
This Green function can be constructed from a complete set of normalized eigenfunctions (see e.g. \cite{Titchmarsh}).
The eigenfunctions of the operator $L$ can be written in terms of Bessel functions of the first and second kinds. They read:
\be\label{contModes} \sqrt{q z} J_2(qz) e^{i p_\mu x^\mu} \qquad ; \qquad \sqrt{q z} Y_2(qz) e^{i p_\mu x^\mu}\ee
The functions \eqref{contModes} are delta-function normalizable as $z$ goes to infinity, i.e. close to the horizon. In this region they can be approximated by trigonometric functions:
\be\label{largeArgBessel} \sqrt{q z} J_{2}(qz) \approx \sqrt{\frac{2}{\pi}} \cos \left( qz - \frac{5\pi}{4} \right) 
\qquad ; \qquad
\sqrt{q z} Y_{2}(qz) \approx \sqrt{\frac{2}{\pi}} \sin \left( qz - \frac{5\pi}{4} \right)
\ee
Near the AdS boundary at $z=0$ the functions $\sqrt{q z} Y_2(qz)$ are not normalizable%
\footnote{Depending on the geometry far away from the stack of D3-branes it may be the case that two linearly independent sets of graviton wave-functions are (delta-function) normalizable, as it happens in flat spacetime. This would not qualitatively modify our main results, but merely add factors of two in the relevant formulas.}.
We deduce that the Green function is:
\be\label{Gnc} G(x^a,{x'}^a) = - \int \frac{d^4 p}{(2\pi)^4} e^{i p_\mu (x^\mu-x'_\mu)} \int_0^\infty dq  \frac{\sqrt{q z} J_2(qz) \sqrt{q z'} J_2(qz')}{q^2 + p_\mu p^\mu} \ee
To study time-independent configurations it is convenient to introduce the time-integrated Green function:
\be\label{Vnc} V(z,\vec x,z',\vec x') = \int_{- \infty}^\infty dt' G_{nc}(x^a,{x'}^a) 
= -\int_0^\infty dq \frac{1}{4\pi} \left(\sqrt{q z} J_2(qz) \sqrt{q z'} J_2(qz')\right) \frac{e^{-qr}}{r} \ee
where we introduced $r=|\vec x - \vec x'|$.
In a Newtonian regime, the time-integrated Green-function is proportional to the potential energy between two static point-like sources. 
This potential energy can be understood from a four-dimensional perspective as a sum of contributions from the Kaluza-Klein modes of the bulk graviton field.
Let us briefly discuss the asymptotic $r$-dependence of the time-integrated Green-function, for $z=z'$. At large distances, i.e. for $r \gg z$, we have:
\be\label{potr7} V(z,\vec x,z,\vec x') \propto \frac{1}{r^7} \ee
This is the same behavior as for the potential energy associated with a massless interaction in ten-dimensional flat spacetime.
On the other-hand, at small distances ($r \ll z$) we have: 
\be\label{potr2} V(z,\vec x,z,\vec x') \propto \frac{1}{r^2} \ee
This is the same behavior as for a massless interaction in five-dimensional flat spacetime. This is to be expected since at small geodesic distances with respect to the $AdS_5$ radius, the effect of the curvature becomes negligible.


\section{Gravitational backreaction of open strings}\label{backreaction}

In this section we consider a static open string ending on a stack of D3-branes. We will compute the linear backreaction of this string on spacetime geometry using the results of the previous section. Similar computations can be found in \cite{Giddings:2000mu}\cite{Garriga:1999yh}.


\subsection{Nambu-Goto open string in Poincar\'e $AdS_5$}\label{NGString}

We consider a classical string living in the Poincar\'e patch of $AdS_5$. 
The string is described by the Nambu-Goto action :
\be S_{NG} = -\frac{1}{2\pi \alpha'}  \int d \sigma d \tau \sqrt{-\det_{i,j=\sigma,\tau}(g_{ab} \p_i x^a \p_j x^b)} \ee

We want to study open strings attached to D3-branes, so we will consider solutions of the Nambu-Goto equations of motion that extend from $z=z_0$ up to the Poincar\'e horizon at $z=\infty$.
We can think of such a string as being stretched between a large stack of D3-branes at $z=\infty$, and a parallel probe D3-brane located at $z=z_0$. 

\begin{figure}
\centering
\includegraphics[scale=0.65]{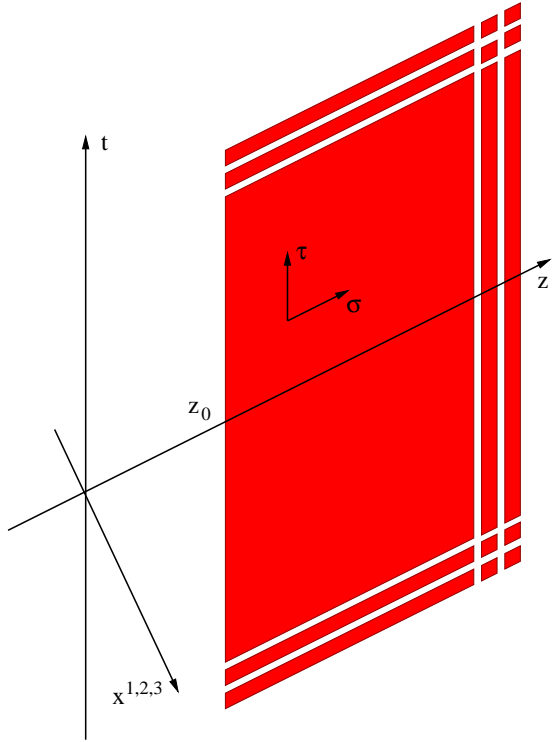}
\caption{The classical string defined in \eqref{classicalString}. It extends from $z=z_0$ up to a stack of D3-branes located at $z=\infty$. \label{string}}
\end{figure}

For simplicity we study static configurations only. 
It is straightforward to check that the following string satisfies the Nambu-Goto equations of motion:
\be\label{classicalString}
t=\tau \qquad ; \qquad
x^{1,2,3}=cst \qquad ; \qquad
z=\sigma \ee
with $\sigma $ running over the interval $[z_0, \infty [$. This string is represented in Figure \ref{string}.
On this solution the Nambu-Goto action evaluates to:
\be S_{NG} = -\frac{1}{2\pi \alpha'} \frac{R^2}{z_0} \int d\tau = -m_0 \int d\tau  \ee
where $m_0$ is the inertial mass of the string:
\be\label{massString} m_0 = \frac{1}{2\pi \alpha'} \frac{R^2}{z_0} \ee
%
We define the length of the string \eqref{classicalString} as the proper length of the (one-dimensional) intersection between the worldsheet and a spacelike hypersurface.
A natural slicing of spacetime is given by the hypersurfaces normal to the timelike Killing vector of the D3-branes background: these are the constant-$t$ hypersurfaces.
With this natural definition the length of the string is infinite:
\be \int_{z_0<z<\infty , \ t=cst} ds = \infty \ee
The effective string tension is redshifted towards the horizon, which renders the mass of the string finite even if its length is infinite. 

If both ends of a static string are attached to the stack of D3-branes at $z=\infty$, the Nambu-Goto equations of motion require that the entire worldsheet be located on the horizon. 
It is reasonable to expect qualitative agreement between the gravitational behavior of the string defined in \eqref{classicalString}, and an oscillating string of energy $m_0$ with both endpoints attached to the stack of D3-branes.


\subsubsection*{The stress-energy tensor}

The Nambu-Goto action can be written as the integral over spacetime of a Lagrangian density:
\be S_{NG} = \int d^5 x \mathcal{L}_{NG}
= - \int d^{5}x  \frac{1}{2\pi \alpha'} \int d \sigma d \tau \sqrt{-det_{i,j}(g_{a b} \p_i x^a \p_j x^b)}\ \delta^{(5)}(x^a - x^a(\tau,\sigma))  \ee
The string stress-energy tensor is evaluated as: 
\be T_{ab} = -\frac{1}{\sqrt{-g}} \frac{\delta \mathcal{L}_{NG}}{\delta g^{ab}} \ee
For the open string defined in \eqref{classicalString} with $\vec{x}=0$, the non-zero components of the stress-energy tensor are:
\be T_{tt} = -T_{zz} =  \frac{1}{2\pi \alpha'} \frac{z}{R} \theta(z-z_0)   \delta^{(3)}(\vec x) \ee
Since spacetime admits a timelike Killing vector $\xi^a = \p_t$, we can construct a covariantly constant mass-energy current ${T^a}_b \xi^b$. The mass of the string $m_0$ can be recovered as the flux of this current through a spacelike hypersurface.


\subsection{Computation of the linear backreaction}

We have now gathered all the ingredients needed to compute the linear gravitational backreaction of the string defined in \eqref{classicalString}. 
First we compute the various terms appearing on the right-hand side of equation \eqref{linEqPhimunu}.
From equation \eqref{linEqH} we deduce $\p_z h$:
\be \p_z h = \frac{G_N^{(5)}}{2\pi \alpha'} \frac{z^2}{3R} \theta(z-z_0) \delta^{(3)}(\vec x) \ee
and from equation \eqref{linEqH2} we deduce $ \p^\nu \bar h_{\mu \nu}$ :
\be \p^\nu \bar h_{\mu \nu}  =
 \frac{G_N^{(5)}}{2\pi \alpha' } \frac{z^3}{18 R} \p_{\mu} \delta^{(3)}(\vec x) \theta(z-z_0) 
\ee 
The integration constants were chosen such that $\p_z h = \p^\nu \bar h_{\mu \nu}=0$ away from the string.

We can now integrate equation \eqref{linEqPhimunu} to obtain $\phi_{\mu \nu}$:
\begin{align}
& \phi_{\mu \nu}(x)  = \int d^5 x' G(x,x') [L \phi_{\mu \nu}](x') \cr
&= \int d^5 x' G(x,x') \left(\frac{z'}{R}\right)^{-\frac{3}{2}} \cr
& \qquad  
\times \left\{ \frac{G_N^{(5)}}{2\pi \alpha' } \frac{z'}{R} \left( \frac{{z'}^2}{18}(2\p_\mu \p_\nu - \eta_{\mu \nu} \p^\rho \p_\rho )- \frac{1}{6}\eta_{\mu \nu} \right) \delta^{(3)}(\vec x') \theta(z'-z_0) - G_N^{(5)} T_{\mu \nu} \right \}
\end{align}
Let us compute explicitly $\phi_{tt}$:
\begin{align}
 &\phi_{t t}(x)  
= \int d^5 x' G(x,x') \left(\frac{z'}{R}\right)^{-\frac{3}{2}} 
\left\{ \frac{G_N^{(5)}}{2\pi \alpha' } \frac{z'}{R} \left( \frac{{z'}^2}{18} \p^\rho \p_\rho +\frac{1}{6}-1 \right) \delta^{(3)}(\vec x') \theta(z'-z_0)  \right \} \cr
&= \frac{G_N^{(5)}}{2\pi \alpha' } \int d^5 x' \left( \frac{{z'}^2}{18} \left( -\p^2_{z'} + \frac{15}{4{z'}^2} \right) - \frac{5}{6} \right) G(x,x') \left(\frac{z'}{R}\right)^{-\frac{1}{2}} \delta^{(3)}(\vec x') \theta(z'-z_0)  \cr
&= \frac{G_N^{(5)}}{2\pi \alpha' } \int d^5 x' \left( \frac{-2}{3} \right) G(x,x') \left(\frac{z'}{R}\right)^{-\frac{1}{2}} \delta^{(3)}(\vec x') \theta(z'-z_0)
\end{align}
From the first line to the second we integrated by parts the derivatives with respect to the Minkowski coordinates, and used equation \eqref{defG}. From the second line to the third we integrated by parts the $z'$ derivatives.
Next we perform the integral over the Minkowski coordinates:
\begin{align}\label{phittstep2}
\phi_{tt}(x) &= -\frac{2}{3} \frac{G_N^{(5)}}{2\pi \alpha' } \int dz' V(z,\vec x,z',0)  \left(\frac{z'}{R}\right)^{-\frac{1}{2}} \theta(z'-z_0) \cr
& = \frac{2}{3} \frac{G_N^{(5)}}{2\pi \alpha' } \int dz' \int_0^\infty dq \frac{1}{4\pi} \left(\sqrt{q z} J_2(qz) \sqrt{q z'} J_2(qz'))\right) \frac{e^{-qr}}{r} \left(\frac{z'}{R}\right)^{-\frac{1}{2}} \theta(z'-z_0) \cr
& =  \frac{G_N^{(5)} \sqrt{R}}{12\pi^2 \alpha' } \int_0^\infty dq q \sqrt{z} J_2(qz)\frac{e^{-qr}}{r} \int_{z_0}^\infty dz'J_2(qz')
\end{align}
From the first line to the second we used  expression \eqref{Vnc} for the time-integrated Green function. We introduced $r=|\vec x|$ which is the euclidean distance away from the string.
To proceed further we have to evaluate an integral over the Bessel function $J_2$.
A primitive of this function can be written in terms of the following generalized hypergeometric function:
\be\label{1F2} {}_1F_2\left(\left\{\frac{3}{2} \right\}, \left\{ \frac{5}{2},3 \right\},u\right) = \sum_{n=0}^\infty \frac{6}{3+2n} \frac{ u^{n}}{ (n+2)! n!} \ee
At large negative argument this function behaves like:
\be\label{1F2inf} {}_1F_2\left(\left\{\frac{3}{2} \right\}, \left\{ \frac{5}{2},3 \right\},-u\right) \stackrel{u \to  \infty}{=} \frac{3}{u^{\frac{3}{2}}} + ... \ee
The integral relevant for our computation is:
\be \int_{z_0}^\infty dz'J_2(qz') = \frac{1}{q} - \frac{z_0^3 q^2}{24} {}_1F_2\left(\left\{\frac{3}{2} \right\}, \left\{ \frac{5}{2},3 \right\},-\frac{z_0^2 q^2}{4}\right) \ee
The two terms in the previous equation give two contributions to the computation \eqref{phittstep2} of $\phi_{tt}$, that we denote respectively $\phi_{tt}^{(\infty)}$ and $\phi_{tt}^{(z_0)}$.
The first one can be explicitly evaluated as:
\be\label{phiInfExact} \phi_{tt}^{(\infty)}(x) =  \frac{G_N^{(5)} \sqrt{R}}{12\pi^2 \alpha' } \int_0^\infty dq \sqrt{z} J_2(qz)\frac{e^{-qr}}{r} 
 = \frac{G_N^{(5)} \sqrt{R}}{12\pi^2 \alpha' }\frac{\left(1 -  \sqrt{1+\frac{z^2}{r^2}}\right)^2}{z^{\frac{3}{2}}\sqrt{1+\frac{z^2}{r^2}}} \ee

\begin{figure}
\centering
\includegraphics[scale=0.65]{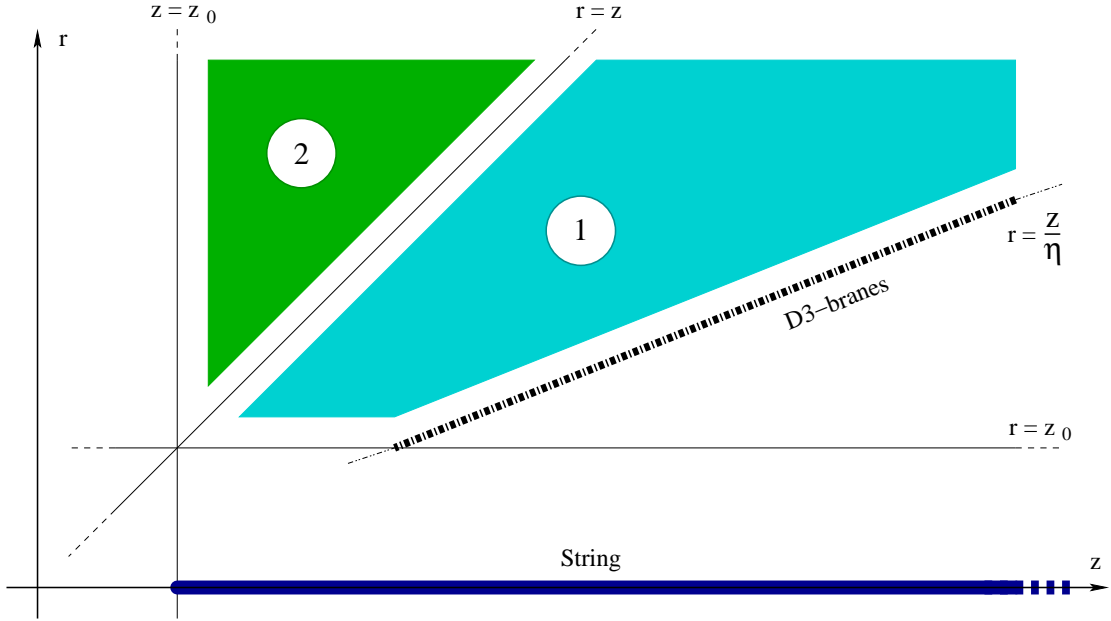}
\caption{Backreaction of an open string in $AdS_5$. The string lies at $r=0$ and extends up to $z=\infty$. 
In region 1 the decay of the gravitational field away from the string is slow. \label{geoNC}}
\end{figure}

From now on we focus on the region of spacetime where $z \gg r \gg z_0$, which will be the most interesting one for our purposes. In this domain, which is denoted as region 1 in Figure \ref{geoNC}, $\phi_{tt}^{(\infty)}$ simplifies to:
\be\label{phinc} \phi_{tt}^{(\infty)}(x) =  \frac{G_N^{(5)} \sqrt{R}}{12\pi^2 \alpha' }\frac{1}{r \sqrt{z}} \ee
Notice that this result depends only on the large-argument behavior of the eigenfunctions of the operator $L$, given in equation \eqref{largeArgBessel}. In other words it relies only on the way the graviton wave-functions behave close to the stack of D3-branes.
It is not sensitive to the geometry far away from the stack of D3-branes.

Now let us consider the second contribution to $\phi_{tt}$, namely $\phi_{tt}^{(z_0)}$.
We can evaluate this contribution thanks to a small-argument expansion of the generalized hypergeometric function, keeping only the first term in the infinite sum \eqref{1F2}. We find that $\phi_{tt}^{(z_0)}$  gives a small correction to the previous result: 
\begin{align}  \frac{\phi_{tt}^{(z_0)}}{\phi_{tt}^{(\infty)}} = \mathcal{O} \left(\frac{z_0}{z}\right)^3 
\end{align}
%
Let us briefly discuss this intermediate result.
The first contribution $\phi_{tt}^{(\infty)}$ is the result associated to an infinite string. 
The second term $\phi_{tt}^{(z_0)}$ can be understood as a correction due to the fact that the string does not extend up to the boundary at $z=0$, but ends up at $z=z_0$. 
The previous equation states that this correction is irrelevant in the region  $z \gg r \gg z_0$, and the string behaves like an infinite string. This is actually expected: in this region the geodesic distance away from the string is small (even if $r$ is large), so the string appears as an infinite string in flat spacetime. The flat-space analysis of appendix \ref{Newton} then predicts that the gravitational backreaction goes like $1/r$. This is in agreement with the computation we just performed.
Let us also mention that a point-like object would produce a different backreaction, which would go like $1/r^2$ or $1/r^7$ depending on whether the geodesic distance away from this object is respectively smaller or larger than the $AdS$ radius (see equations \eqref{potr7} and \eqref{potr2}).

From equations \eqref{phinc}, \eqref{phi(h)}, \eqref{defBarH} and using the fact that $\bar h$ is traceless, it is straightforward to deduce the linear backreaction of the string in the region $z \gg r \gg z_0$:
\be\label{hnc} h_{tt} = \frac{G_N^{(5)}}{12 \pi^2 \alpha'  R} \frac{z}{r} \ee
The other components of the metric perturbation can be computed along the same lines, or deduced from $h_{tt}$ using symmetry and the gauge choices we made.
Let us stress again that this result is independent of the geometry far away from the D3-branes.

Linear perturbation theory breaks down as the metric perturbation becomes of order one. 
Actually the curve defined by $h_{tt}=1$ gives an estimation of the location of the horizon created by the string.
Notice that the intensity of the gravitational field at a given $r$ increases with $z$:
 the redshift of the effective string tension towards the horizon is overwhelmed by the decrease of the geodesic distance away from the string.

In the other regions of spacetime the linear backreaction typically comes with a larger (negative) power of $r$.
For instance in the domain where $r \gg z \gg z_0 $ (which is denoted as region 2 in Figure \ref{geoNC}) equation \eqref{phiInfExact} gives:
\be\label{phincsub} \phi_{tt}^{(\infty)}(x) = \frac{G_N^{(5)}\sqrt{R}}{12\pi^2 \alpha'}\frac{1}{4} \frac{z^{\frac{5}{2}}}{r^4} \ee
As previously $\phi_{tt}^{(z_0)}$ gives a small correction:
\begin{align} 
\frac{\phi_{tt}^{(z_0)}}{\phi_{tt}^{(\infty)}} = \mathcal{O} \left(\frac{z_0}{r}\right)^3
\end{align}
These results follow from a small-argument expansion of the Bessel functions. Thus they depend on the geometry far away from the stack of D3-branes.

We will not try to evaluate the metric perturbation outside the domain where $r,z \gg z_0$ since it will not be needed for the computation we have in mind, namely the evaluation of the gravitational attraction between two open strings.
Moreover the result would be sensitive to the details of the geometry far away from the D3-branes.


\subsection{Deformation of the stack of D3-branes}
\label{DBI}

Before we added the open string, the stack of D3-branes was located on the Poincar\'e horizon at $z=\infty$. 
The string induces a deformation of the stack of D3-branes, which can be computed using the DBI action.
This analysis was performed in \cite{Callan:1997kz} for a single D3-brane.
The solution to the DBI equations of motion corresponding to a semi-infinite fundamental string ending on a single D3-brane reads:
\be\label{rDBI1} r^{(N=1)}(\rho) = \frac{g_s \alpha' \pi}{\rho} \ee
%
where $r$ is the euclidean distance away from the string along the D3-brane, and $\rho=\frac{R^2}{z}$ is the radial coordinate away from the D3-brane (cf equation \eqref{ds2spacetime} and Figure \ref{2strings}). We identified $\rho$ with the excited scalar field living on the D3-brane worldvolume, in agreement with the analysis of \cite{Gauntlett:1999xz} \cite{Ghoroku:1999bc} where it is shown that the backreaction of D3-branes essentially does not modify the flat-space result derived in \cite{Callan:1997kz}.

A generalization of this solution for a fundamental string ending on a stack of $N$ D3-branes can be obtained by assuming that the $N$ D3-branes are all equally deformed.
Then the stack of $N$ D3-branes behaves effectively like a single D3-brane for which the tension has been multiplied by $N$.
It follows that the solution we are looking for is the same as $\eqref{rDBI1}$, up to a normalization. We can fix this normalization for instance by demanding that the energy of the solution be equal to the string tension times the length of the string. We find:
\be\label{rDBIN} r^{(N)}(\rho) = \frac{g_s \alpha' \pi}{N\rho} \ee
The assumption that all D3-branes are deformed in the same way means that the open string endpoint sources the diagonal $U(1)$ in the $U(N)$ gauge theory living on the D3-branes.
There are other solutions to the DBI equations of motions for which this assumption is not satisfied.
We will consider only the solution \eqref{rDBIN}, which will be sufficient for our purposes.
Let us also point out that the radius of the string horizon estimated from equation \eqref{hnc} by setting $h_{tt}=1$ matches with the DBI radius \eqref{rDBIN}, up to a numerical factor.
This suggests that the horizon created by the string can be identified with the position of the (deformed) stack of D3-branes.
A similar interpretation can be found in \cite{Lunin:2007mj}.

In terms of the $z$ coordinate, we obtain that the deformed stack of D3-branes is located along the curve $z_{D3}(r)$ given by:
\be\label{zD3} z_{D3} = \eta r \ee
where $\eta$ is a dimensionless parameter.
Using equation \eqref{radius} we can estimate $\eta$ in terms of $g_s$ and $N$:
\be\label{eta} \eta 
\approx \sqrt{\frac{N^3}{g_s}} \ee
where we discarded a numerical factor that is not relevant for our purposes.
We notice that as soon as the string coupling is small, the parameter $\eta$ is large.
%


\section{Gravitational interactions between open strings}\label{potential}

Let us now consider two static open strings that are attached to the same stack of D3-branes.
In this section we will estimate the gravitational potential energy between these two strings, by evaluating the Nambu-Goto action for one  string in the gravitational field created by the other one.
We do not have the ambition to provide an exact computation, thus we will not keep track of the numerical factors.

Let us mention at that point that we are doing a computation in a non-supersymmetric theory. 
The static configuration of fundamental strings and D3-branes that we are considering is typically a BPS configuration once embedded into superstring theory.
In that case no net force is expected between the strings.
This means that the gravitational interaction is exactly canceled by other forces, presumably associated to the Kalb-Ramond and dilaton fields.
Nevertheless the gravitational force we evaluate has a physical manifestation in non-supersymmetric circumstances.


\subsection{Evaluation of the gravitational potential energy}\label{potentialnc}

\begin{figure}
\centering
\includegraphics[scale=1]{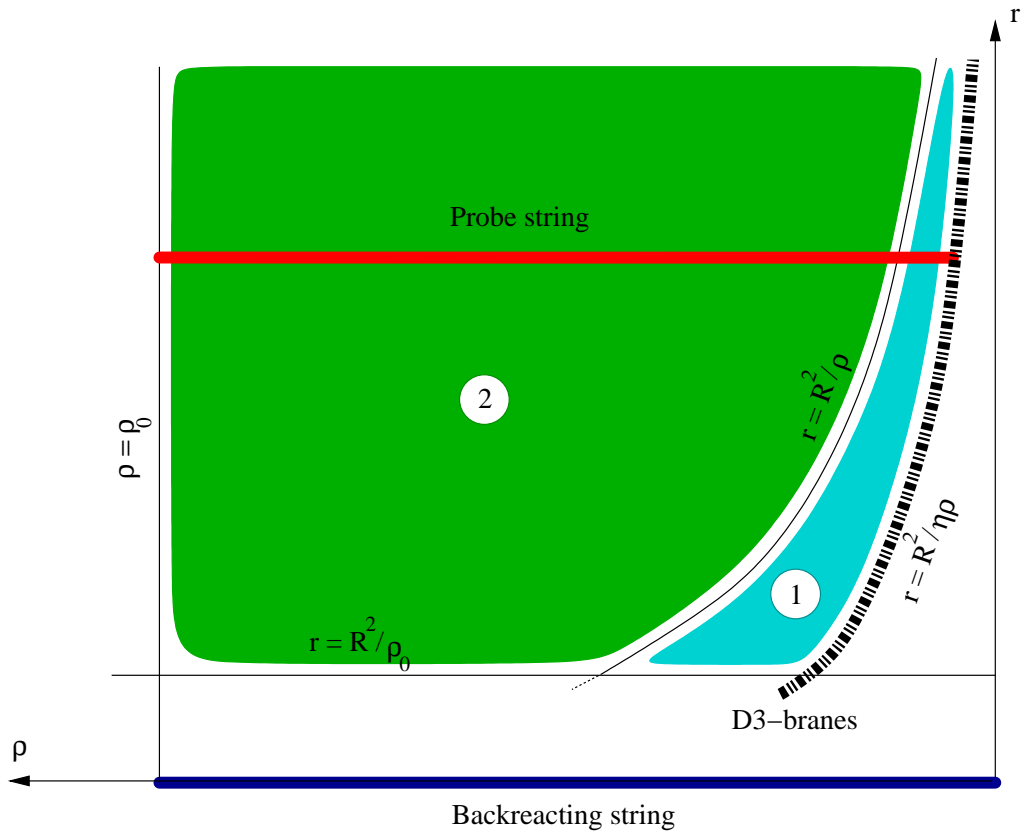}
\caption{Evaluation of the gravitational attraction induced by the backreacting string upon the probe string.
The gravitational potential energy is dominated by a contribution coming from region 1, localized near the endpoint of the probe string.
\label{2strings}}
\end{figure}

The strings are parametrized as in \eqref{classicalString}. We denote by $r$ the euclidean distance between the two strings, as measured by an observer living on the D3-branes.
We are interested in the macroscopic behavior of the gravitational interaction, hence we assume that the distance $r$ is large (in particular $r$ is much larger than $z_0$). 
%
Up to first order in the metric perturbation, the Nambu-Goto action for a probe string placed in the gravitational field created by a backreacting string (cf Figure \ref{2strings}) reads:
\beq S_{NG}\label{SNG2strings}  
& \approx & - \frac{1}{2\pi \alpha'} \int d t \int_{z_0}^{z_{D3}} d z \frac{R^2}{z^2} \left(1-\frac{h_{tt}}{2} \right) \eeq
The integral over $z$ extends from the probe string endpoint at $z=z_0$ up to the point where the probe string touches the stack of D3-branes at $z=z_{D3}$, that we evaluated in equation \eqref{zD3}.

We are interested in the large distance behavior of the gravitational potential energy between the strings. 
In the evaluation of the Nambu-Goto action \eqref{SNG2strings} the integral over $z$ is dominated by the region $r<z<z_{D3}$ where the perturbation of the metric is of order $r^{-1}$:
\beq S_{NG} &\approx & -  \int d\tau \left(m_0-\frac{G_N^{(5)} }{\alpha'^2  r} \int_{r}^{z_{D3}} dz \frac{R}{z} \right)
\eeq
The second term in the previous expression gives the gravitational potential energy between the two static strings. 
The integral appearing in this term is the proper length of the piece of the probe string contained in the region $r<z<z_{D3}$.
It turns out to be independent of $r$:
\be\label{propLengthR1} \int_{r}^{z_{D3}} dz \frac{R}{z} = R \log \eta \ee
where $\eta$ is the dimensionless parameter introduced in \eqref{zD3}.
We deduce an estimation for the gravitational potential energy:
\be\label{Enc} E(r) \approx - G_N^{(5)} \frac{1}{\alpha'} \frac{ R \log \eta }{\alpha' } \frac{1}{r} \ee
%


\subsection{Interpretation}\label{interpretation}

Let us now discuss the interpretation of the result \eqref{Enc} we just obtained.
First let us stress that the dominant contribution to the gravitational interaction between the strings is localized near the endpoints of the strings, as can be seen in Figure \ref{2strings}.
Accordingly the potential \eqref{Enc} does not depend on the inertial masses of the strings involved.
Indeed the inertial mass of the probe string in Figure \ref{2strings} is almost entirely located in region 2, that provides a subleading contribution to the gravitational potential.

The second important point is that the potential energy goes like $1/r$ at 
arbitrary
large $r$ despite the fact that gravity lives in more than four dimensions. 
This behavior is very different from what we would obtain for point-particles. 
Indeed the Newtonian potential between two point-particles goes like $1/r^7$ in $AdS_5$ (cf equation \eqref{potr7}), and it goes like $1/r^2$ in five-dimensional flat space.
Notice that this result is also qualitatively different from the gravitational potential energy between strings of finite energy in flat spacetime (see appendix \ref{Newton}): as soon as the distance between such strings is much larger than their size, they would behave like point-particles.
The potential \eqref{Enc} has the same large-$r$ behavior as for point particles in four-dimensional flat spacetime, or equivalently as for infinite parallel strings in five-dimensional flat spacetime (cf Appendix \ref{Newton}).

It is interesting to compare our result \eqref{Enc} with the gravitational potential \eqref{potClose5D} for a point mass close to a one-dimensional extended object.
The open strings potential energy \eqref{Enc} matches with the Newtonian potential between an infinite string of tension $1/\alpha'$ and a mass ${ R \log \eta }/{\alpha' }$ in five-dimensional flat space.
This mass is the product between the probe string tension and the proper length \eqref{propLengthR1}.

\begin{figure}
\centering
\includegraphics[scale=0.95]{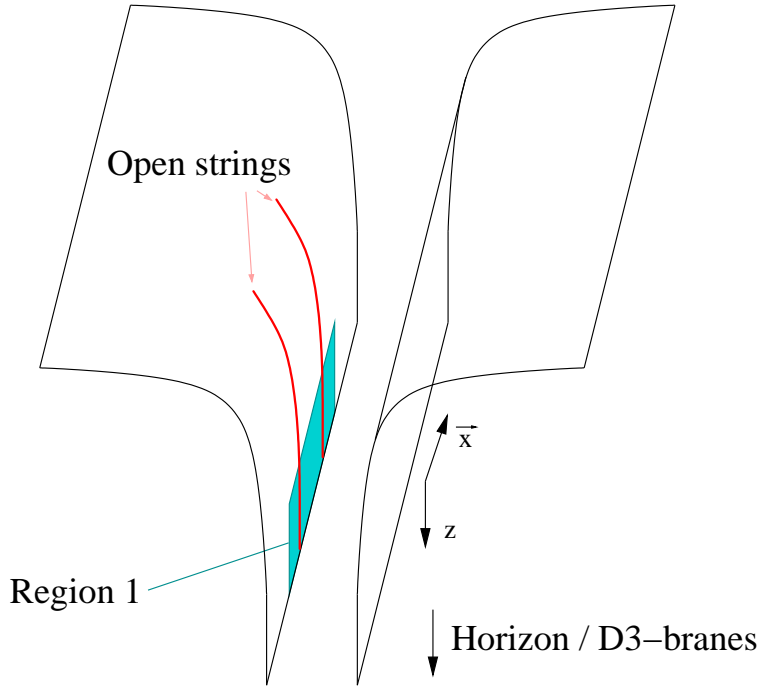}
\caption{The D3-branes' backreaction creates an infinite throat. In region 1 the geodesic distance between the open strings is small compared to the curvature radius, and the open strings interact like infinite strings in flat space. This region provide the leading contribution to the potential energy \eqref{Enc}.
\label{interpretationFig}}
\end{figure}

We can now provide an interpretation for the result \eqref{Enc}, which is illustrated in Figure \ref{interpretationFig}.
Even if the two strings are widely separated, the geodesic distance between points located near the Poincar\'e horizon is very small. 
At distances that are much smaller than the $AdS$ radius the influence of the curvature is negligible and spacetime is essentially flat.
Close to the horizon the strings appear as being infinitely extended along the radial direction, and thus interact like infinite strings in flat space.
Notice that the redshift of the strings' tension near the horizon is compensated by the decrease of the geodesic distance between the strings.

The usual expectation when dealing with extended objects of finite energy is that a point-particle approximation is reliable as soon as the distance away from the objects is large enough.
For open strings attached to backreacting D3-branes, our results show that this is not the case%
\footnote{This is true as far as bulk interactions are concerned. We are not considering interactions mediated by degrees of freedom living on the branes.}.
Indeed the classical open strings we are considering have a finite energy, which can be arbitrarily small.
But from the point of view of gravity they behave very differently than point-particles.
This holds even if the distance $r$ separating the strings (as measured by an observer living on the branes) is arbitrarily large.
The reason is essentially that the backreaction of the D3-branes stretches the open strings up to an infinite length, and decreases the geodesic distance between the strings as we approach the D3-branes' horizon.
%
Thus we identified large-distance manifestation of the microscopic differences between point-particles and strings.
We will give further comments on this point in section \ref{discussion}.

\section{Higher dimensional spacetimes}\label{generalizationDim}

Let us now discuss the influence of the compact transverse space that typically appears in the spacetime metric \eqref{ds2spacetime} in the neighborhood of D3-branes.
Since our goal is once again to evaluate the gravitational energy potential between two open strings, we will repeat the steps we followed previously in a new set-up.
To avoid too much repetition we will focus on the differences with the previous analysis.

The discussion of section \ref{interpretation} already gives some clues about what the result will be.
The enhancement of the gravitational potential energy between open strings \eqref{Enc} is due to the fact that the geodesic distance between the strings is small near the horizon.
If the strings are widely separated along an additional transverse compact manifold, we can expect that the dominant gravitational interaction localized near the string endpoints will fade away.
On the other hand if the strings are close enough along the transverse compact manifold, we should recover a stringy enhancement of the gravitational attraction.

We will now make this argument more precise.
For the sake of simplicity we assume that the compact space is a circle of radius $\tilde{R}$, but the generalization to higher-dimensional compact spaces is rather straightforward (the result will be given in section \ref{summary}).
Let us also mention that $AdS_5 \times S^1$ is the near-horizon geometry of D3-branes in non-critical six-dimensional backgrounds \cite{Klebanov:2004ya}.

\subsubsection*{Green function}

We want to construct a Green function that allows for the integration of the wave equation in $AdS_5 \times S^1$.
We perform the same field redefinition as in \eqref{phi(h)}. We obtain:
\be \Box h = \left(\frac{z}{R}\right)^{\frac{7}{2}} \tilde{L} \phi \ee
The linear differential operator $\tilde{L}$ is a six-dimensional generalization of the operator $L$ introduced in \eqref{opL}:
\be \tilde{L} = \p_z^2 - \frac{15}{4z^2} + \eta^{\mu \nu} \p_\mu \p_\nu + \frac{R^2}{\tilde{R}^2} \frac{ \p_\theta^2}{z^2} \ee
where the angular variable $\theta$ parametrizes the circle.
The delta-function normalizable eigenfunctions of the operator $\tilde L$ are:
\be \sqrt{qz} J_{\nu_n}(qz) e^{i p_\mu x^\mu} e^{ i n \theta} \ee
with
\be \nu_n = \sqrt{4+n^2 \frac{R^2}{\tilde{R}^2}} \ee
The Green function is then:
\be\label{} \tilde{G}(x^a,{x'}^a) = -  \int \frac{d^4 p}{(2\pi)^4} e^{i p_\mu (x^\mu-x'_\mu)} \sum_{n=-\infty}^\infty e^{in(\theta-\theta')} \int_0^\infty dq  \frac{\sqrt{q z} J_{\nu_n}(qz) \sqrt{q z'} J_{\nu_n}(qz')}{q^2 + p_\mu p^\mu} \ee
and the time-integrated Green function reads:
\be\label{} \tilde{V}(z,\vec x,\theta,z',\vec x',\theta') 
= - \sum_{n=-\infty}^\infty \int_0^\infty dq \frac{1}{4\pi} \left(\sqrt{q z} J_{\nu_n}(qz) \sqrt{q z'} J_{\nu_n}(qz')\right) e^{in(\theta-\theta')} \frac{e^{-qr}}{r} \ee
The behavior of this potential in the region $r \ll z,z'$ depends on the large-argument behavior of the Bessel functions, which is given by:
\be \sqrt{x} J_{\nu_n}(x) \approx \sqrt{\frac{2}{\pi}} \cos \left( x - \nu_n \frac{\pi}{2} - \frac{\pi}{4} \right) \ee

\subsubsection*{Linear backreaction of an open string}

\begin{figure}
\centering
\includegraphics[scale=0.6]{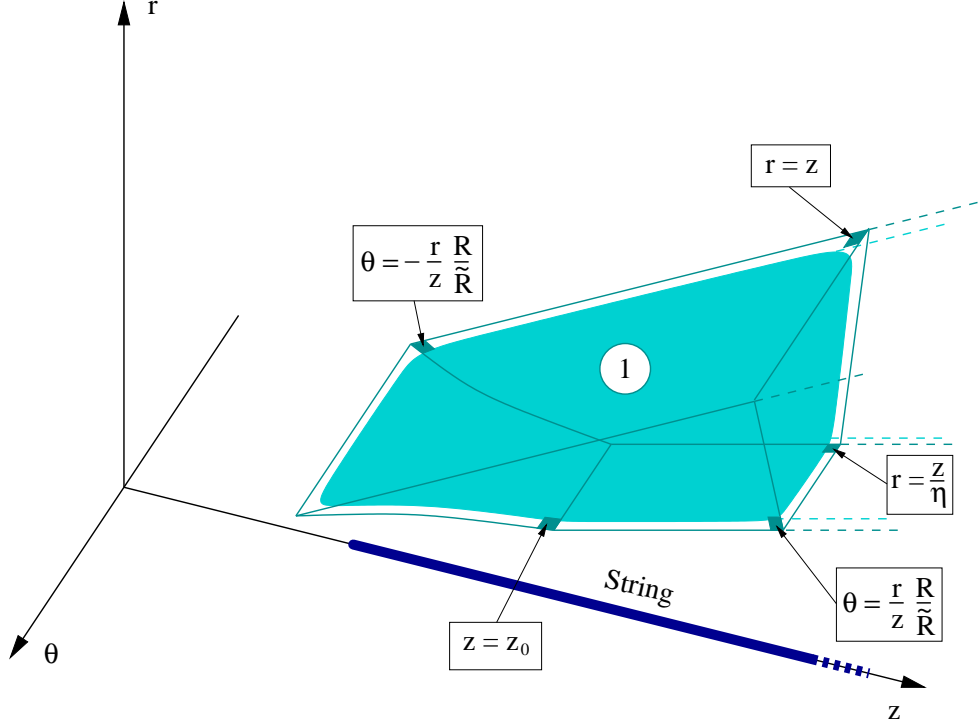}
\caption{Backreaction of an open string in $AdS_5 \times S^1$.
In region 1 the decay of the gravitational field away from the string is slow.
The different equations define the surfaces that surround this region.
\label{geo3D}}
\end{figure}

We consider a classical string extending up to the Poincar\'e horizon, defined by:
\be
t=\tau \qquad ; \qquad
x^{1,2,3}=0 \qquad ; \qquad
z=\sigma
\qquad ; \qquad
\theta = 0 \ee
We compute the linear gravitational backreaction of this string.
We obtain for $\phi_{tt}$ (cf equation \eqref{phittstep2}):
\begin{align}
\phi_{tt}(x) &\approx -\frac{G_N^{(6)}}{ \alpha' \tilde{R}} \int dz' d \theta' \tilde{V}(z,\vec x,\theta,z',0,\theta')  \left(\frac{z'}{R}\right)^{-\frac{1}{2}} \theta(z'-z_0) \delta(\theta') \cr
& \approx \frac{G_N^{(6)}\sqrt{R}}{ \alpha' \tilde{R}} \sum_{n=-\infty}^{\infty}  \int_0^\infty dq q \sqrt{z} J_{\nu_n}(qz) e^{i n \theta} \frac{e^{-qr}}{r} \int_{z_0}^\infty dz' J_{\nu_n}(qz')
\end{align}
The integral of the Bessel function can be estimated as:
\be \int_{z_0}^\infty dz' J_{\nu_n}(qz') = \frac{1}{q} - ... \ee
where the ellipses contain a term which depends on $z_0$. This term gives a subleading correction to the linear backreaction in the region $z \gg r \gg z_0$.
In this particular region we obtain:
\begin{align}
\phi_{tt}(x) &\approx \frac{G_N^{(6)}\sqrt{R}}{\alpha'\tilde{R} } \sum_{n=-\infty}^{\infty} e^{i n \theta} \int_0^\infty dq  \sqrt{z} J_{\nu_n}(qz) \frac{e^{-qr}}{r}
\end{align}
The integral over $q$ can be evaluated in terms of the Gaussian hypergeometric function:
\be \int_0^\infty dq   J_{\nu_n}(qz) e^{-qr} = \frac{z^{\nu_n}}{2^{\nu_n} r^{\nu_n+1}} {}_2 F_1\left( \frac{1 + \nu_n}{2},\frac{2+\nu_n}{2},1+\nu_n,-\frac{z^2}{r^2} \right) \ee
We can expand this function for $z \gg r$:
\be  \frac{z^{\nu_n}}{2^{\nu_n} r^{\nu_n+1}} {}_2 F_1\left( \frac{1 + \nu_n}{2},\frac{2+\nu_n}{2},1+\nu_n,-\frac{z^2}{r^2} \right)
 = \frac{1}{z} - \frac{\nu_n r}{z^2} + \frac{1}{z}\mathcal{O}\left(\frac{r}{z}\right)^2 \ee
We deduce for $\phi_{tt}$:
\begin{align}
\phi_{tt}(x) &\approx \frac{G_N^{(6)}\sqrt{R}}{\alpha' \tilde{R}} \frac{1}{r \sqrt{z}} \sum_{n=-\infty}^{\infty} e^{i n \theta}\left(1 - \frac{\nu_n r}{z}+ \mathcal{O}\left(\frac{r}{z}\right)^2\right)
\end{align}
At large $n$, $\nu_n$ behaves like $|n|R/\tilde{R}$. We recognize%
\footnote{For instance the Lorentzian function with a small scale parameter $\gamma$ satisfies:
\be \int_{-\infty}^{\infty} e^{-in \theta} \frac{1}{\gamma \pi(1+\frac{\theta^2 }{\gamma^2})} = e^{-|n|\gamma } = 1 - |n|\gamma + \mathcal{O} \left(\gamma \right)^2 \ee}
 in the previous equation the Fourier decomposition of a function localized around $\theta=0$, with a width of order $\frac{r}{z} \frac{R}{\tilde{R}}$ and a height of order $\frac{z}{r} \frac{\tilde{R}}{R}$. We approximate this function as:
\be \sum_{n=-\infty}^{\infty} e^{i n \theta}\left(1 - \frac{\nu_n r}{z}+ \mathcal{O}\left(\frac{r}{z}\right)^2\right) 
\approx \left \{ \begin{array}{lll}
\frac{z}{r} \frac{\tilde{R}}{R}  & \textrm{if} & |\theta|< \frac{r}{z} \frac{R}{\tilde{R}} \\ 0 & \textrm{else.}& \end{array} \right. \ee
We observe that the backreaction does not spread in the transverse compact space. It is localized close to the string at $\theta=0$.
In the region where $z \gg r \gg z_0$ and $|\theta|< \frac{r}{z} \frac{R}{\tilde{R}}$ (region 1 in Figure \ref{geo3D}) we deduce for $h_{tt}$:
\be h_{tt} \approx G_N^{(6)} \frac{1}{\alpha' R^2}  \frac{z^2}{r^2}  \ee
The $1/r^2$ behavior of the backreaction was expected: it is the same as for an infinite string in flat six-dimensional spacetime.
The growth of the gravitational field with $z$ (at fixed $r$) is faster than in the five-dimensional case. This is a consequence of the localization of the backreaction along the transverse compact space, which increases with $z$.

In the other regions of spacetime the linear backreaction comes with a larger negative power of $r$.

\subsubsection*{The gravitational potential energy between two open strings}

\begin{figure}
\centering
\includegraphics[scale=0.95]{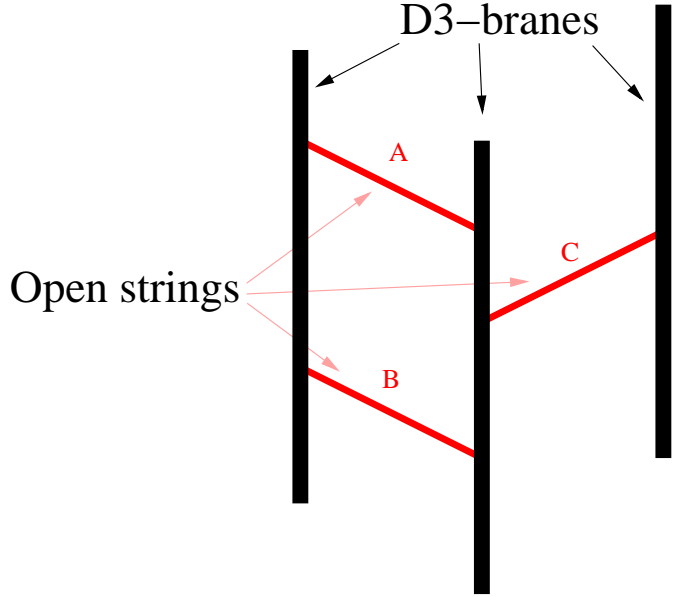}
\caption{The gravitational potential energy between two parallel strings (e.g. A and B) is given in \eqref{E6D}. There is no similar enhancement of the potential energy between strings that extend away from the branes in very different directions (e.g. A and C).
\label{orientation}}
\end{figure}

The gravitational potential energy between two open strings follows from the previous analysis.
First let us consider two strings that are close one to the other in the compact transverse space.
Concretely we assume that the separation $\tilde{R} \theta$ between the two strings along the $S^1$ is much smaller than the $AdS$ radius $R$.
It means that the two strings extend in the same direction away from the D3-branes.
This is always the case in particular for two strings stretching between the same stacks of D3-branes (cf Figure \ref{orientation}).
From the previous discussion we deduce that the gravitational potential energy is dominated by a contribution coming from region 1 in Figure \ref{geo3D}:
\be\label{compE6D} E(r) \approx  \frac{G_N^{(6)}}{\alpha' R^2} \int_r^{\min(z_{D3},z_\theta)} dz \frac{R^2}{z^2} \frac{z^2}{r^2} \ee
where $z_\theta$ is defined as:
\be z_\theta = \frac{r}{\theta} \frac{R}{\tilde{R}} \ee
The upper bound in the $z$ integral in \eqref{compE6D} is chosen such that the integration finishes either when the probe string touches the stack of D3-branes at $z=z_{D3}$, or when the condition $|\theta|< \frac{r}{z} \frac{R}{\tilde{R}}$ fails to be satisfied at $z=z_{\theta}$.
We obtain:
\be\label{E6D}  E(r) \approx  \frac{G_N^{(6)}}{R^2} \frac{R^2}{{\alpha'}^2}\min\left(\eta, \frac{R}{\tilde{R} \theta}\right)  \frac{1 }{r} \ee
In that case we recover the stringy enhancement of the gravitational attraction \eqref{Enc} first obtained in a five-dimensional set-up.

On the other hand if the open strings are broadly separated along the compact transverse space (i.e. for $\tilde{R} \theta \gtrsim R$), or equivalently if they extend in very different directions away from the D3-branes, there is no strong enhancement of the gravitational potential energy due to the extended nature of the strings.

\section{Compactification of the transverse space}\label{generalization}

In the previous sections we studied the gravitational interaction between open strings in non-compact spacetimes. 
Now we will generalize our computations to the case where the space transverse to the stack of D3-branes is compactified.
To this end we will work in the Randall-Sundrum II set-up \cite{Randall:1999vf}: the Poincar\'e patch is truncated to the region $z>R$, and the fields satisfy Neumann boundary conditions along the wall at $z=R$. 
Indeed this set-up can be used as a toy-model for D3-branes localized on a Calabi-Yau compactification of string theory \cite{Verlinde:1999fy}. 
For our purposes, the main difference with the previous cases is that the graviton admits a normalizable collective mode that propagates usual four-dimensional gravity throughout spacetime.


\subsubsection*{Green function}\label{compactGreen}

Linearized Einstein's equations \eqref{linEqs} are unchanged.
To integrate them we have to construct the Green function associated to the linear differential operator $L$ introduced in equation \eqref{opL}.
The Neumann boundary conditions at $z=R$ are satisfied by a linear combination of the eigenfunctions \eqref{contModes}:
\be\label{eigenCompact} \varphi_q(z)e^{i p_\mu x^\mu} = \sqrt{q z} \left(\frac{J_2(qz) Y_1(qR) - Y_2(qz) J_1(qR)}{\sqrt{Y_1^2(qR)+ J_1^2(qR)}} \right) e^{i p_\mu x^\mu}\ee
In addition the operator $L$ admits a normalizable zero mode:
\be\label{eigenCompactColl} \sqrt{2} R z^{-\frac{3}{2}}e^{i p_\mu x^\mu} \ee
The Green function is then:
\be\label{Gc} G_c(x^a,{x'}^a) = - \int \frac{d^4 p}{(2\pi)^4} e^{i p_\mu (x^\mu-x'_\mu)} \left[2R^2\frac{z^{-\frac{3}{2}}{z'}^{-\frac{3}{2}}}{p_\mu p^\mu}  + \int_0^\infty dq  \frac{\varphi_q(z)\varphi_q(z') }{q^2 + p_\mu p^\mu} \right] \ee
where the subscript $c$ stands for compact. The time-integrated Green-function is:
\begin{align}\label{Vc} V_{c}(z,\vec x,z',\vec x')  
& = - \frac{1}{2\pi} \frac{R^2 z^{-\frac{3}{2}}{z'}^{-\frac{3}{2}}}{r} -\int_0^\infty dq \frac{1}{4\pi} \varphi_q(z)\varphi_q(z') \frac{e^{-qr}}{r} \cr
& = V_{0}(z,\vec x,z',\vec x') +  V_{nc}(z,\vec x,z',\vec x')
\end{align}
We wrote the time-integrated Green function as the sum of two terms: $V_{0}(z,\vec x,z',\vec x')$ is the contribution from the normalizable collective-mode of the graviton \eqref{eigenCompactColl},
and $V_{nc}(z,\vec x,z',\vec x')$ is the contribution from the continuum of massive modes \eqref{eigenCompact}. As the notation suggests the latter piece resembles the time-integrated Green function we encountered in the non-compact case.


\subsubsection*{Linear backreaction of an open string}

We now compute the linear backreaction of a classical string extending up to the Poincar\'e horizon, as defined in equation \eqref{classicalString}. 
We obtain for $\phi_{tt}$ (cf equation \eqref{phittstep2}):
\begin{align}\label{phittC1}
\phi_{tt}(x) &= -\frac{2}{3} \frac{G_N^{(5)}}{2\pi \alpha'} \int dz' V_{c}(z,\vec x,z',0)  \left(\frac{z'}{R}\right)^{-\frac{1}{2}} \theta(z'-z_0)
\end{align}
Following equation \eqref{Vc} we decompose $\phi_{tt}$ in two pieces:
\be \phi_{tt} = \phi_{tt}^{(0)} + \phi_{tt}^{(nc)}\ee
First let us compute the contribution $\phi_{tt}^{(0)}$ from the collective mode of the graviton. We obtain:
\be \phi_{tt}^{(0)}(x) = \frac{G_N^{(5)}}{6 \pi^2 \alpha'} \left( \frac{z}{R} \right)^{-\frac{3}{2}} \frac{R}{z_0 r}  \ee
This result is proportional to the inertial mass of the string $m_0$, introduced in equation \eqref{massString}:
\be \phi_{tt}^{(0)}(x) = \frac{G_N^{(5)}}{3 \pi  R}\left( \frac{z}{R} \right)^{-\frac{3}{2}}\frac{m_0}{r}  \ee
Then we consider the second contribution $\phi_{tt}^{(nc)}$ from the continuum of graviton modes.
In the region $z \gg r \gg z_0$ (region 1 in Figure \ref{geoNC}), the linear backreaction 
depends only on the large-argument behavior of the graviton wave-functions.
From equation \eqref{largeArgBessel} we observe that at large argument the compact eigenfunctions \eqref{eigenCompact} behave like the non-compact ones.
It follows that the contribution $\phi_{tt}^{(nc)}$ is essentially identical to the result \eqref{phinc} derived in the non-compact case:
\be \phi_{tt}^{(nc)}(x) = \frac{G_N^{(5)} \sqrt{R}}{12\pi^2 \alpha'}\frac{1}{r \sqrt{z}} \ee
We deduce the full linear backreaction in the region $z \gg r \gg z_0$:
\be\label{httC1} h_{tt}
=  \frac{G_N^{(5)}}{3 \pi  R} \left(m_0 + \frac{z}{4\pi \alpha'} \right) \frac{1}{r} \ee
The first term comes from the collective mode of the graviton and is proportional to the inertial mass of the string.
The second term comes from the continuum of graviton modes and depends on the tension of the string only.
The contribution of the continuum dominates at large $z$.

In other regions of spacetime the continuum of graviton modes only produces a small correction to the collective-mode contribution, at large $r$.
We obtain:
\be\label{httC2}  h_{tt}
=  \frac{G_N^{(5)}}{3 \pi  R} m_0  \frac{1}{r} \ee

As expected the collective mode of the graviton produces a perturbation of the Minkowski metric that is independent of $z$. The resulting spacetime for this contribution only is the five-dimensional black string of \cite{Chamblin:1999by}. This metric receives corrections from the continuum of graviton modes. Near the Poincar\'e horizon the contribution of the continuum actually dominates, thus potentially curing the instability of the black string solution discussed in \cite{Chamblin:1999by}.


\subsubsection*{The gravitational potential energy between two open strings}\label{potentialc}

Next we compute the gravitational potential energy between two open strings.
The collective mode of the graviton induces the expected four-dimensional Newtonian potential between the strings.
The long-range interaction between open-string endpoints mediated by the continuum gives a second contribution to the potential energy:
\be\label{Ec} E_c(r) \approx - \frac{1}{r}\frac{G_N^{(5)}}{R} \left(m_0^2 + 
\frac{R^2}{\alpha'^2} \log\eta \right)\ee
The second term in the parenthesis can be interpreted as the square of a gravitational mass associated to the open string endpoints, which would be independent of the inertial mass of the string. 
The resulting correction to four-dimensional Newton's law is of order $R^2/\alpha'$, which is typically a positive power of the 't Hooft coupling for the gauge theory living on the D3-branes.
This correction induces a violation of the equivalence principle in the four-dimensional low-energy theory.
We will discuss this point further in section \ref{discussion}.


\section{Discussion}\label{discussion}


\subsection{Summary of the results}\label{summary}

Let us summarize the main results we obtained.
We considered open strings attached to D3-branes for which the near-horizon geometry is the product of Poincar\'e $AdS_5$ with a compact space.
We evaluated the gravitational potential energy between two strings at large distances, using a field theory description of gravity.
The result depends on the relative position of the strings on the transverse compact space.
First let us assume that the distance between the strings along the transverse compact space is much smaller than the $AdS$ radius.
We loosely call such strings `parallel strings' since they extend in the same direction away from the D3-branes.
In a non-compact spacetime, the result is:
\be\label{resultnc} E(r) \approx  -\frac{G_N^{(D)}}{R^{D-4}} \frac{R^2\tilde{\eta}}{{\alpha'}^2}  \frac{1 }{r} \ee
where $D$ is the number of dimensions in spacetime.
The dimensionless parameter $\tilde{\eta}$ is a decreasing function of the distance between the strings along the transverse compact space. 
The result \eqref{resultnc} is independent of the geometry far away from the stack of D3-branes.
The inertial masses of the strings do not appear in this formula, but their tension does.
The gravitational interaction is localized near the endpoints of the open strings.
It can be understood as an interaction between infinitely extended strings in flat spacetime.

On the other hand if the distance between the strings along the transverse compact space is of the order of the $AdS$ radius (or equivalently if they extend in different directions away from the D3-branes), there is no strong enhancement of the gravitational attraction due to the extended nature of the strings.

The situation is different if the space transverse to the D3-branes is compactified and
 a collective mode of the graviton propagates Einstein's four-dimensional gravity throughout spacetime. 
The gravitational potential energy between parallel strings is then:
\be\label{resultc} E(r) \approx -\frac{G_N^{(D)}}{V_c} \frac{m_0^2}{r} - \frac{G_N^{(D)}}{R^{D-4}} \frac{R^2\tilde{\eta}}{{\alpha'}^2}  \frac{1 }{r} \ee
where $m_0$ is the inertial mass of the open strings and $V_c$ is the volume of the transverse space.
As previously the second term comes from a short-distance interaction localized near the end-points of the open strings. This term vanishes if the strings are broadly separated along the transverse compact space.


\subsection{About the point-particle and the probe-brane approximations}

It was previously claimed that the results we derived for the gravitational interaction between open strings cannot be reproduced by a point-particle approximation. Here we give more details about this claim.

The gravitational backreaction of an extended object has a different behavior depending on whether the distance away from the object is large or not (with respect to the typical size of the object).
Typically the decay of the backreaction is faster at larger distances, where the object appears as punctual. 
A simple example is given by the string in flat D-dimensional spacetime described in appendix \ref{Newton}. If the distance away from the string is smaller than the string's size, the gravitational backreaction goes like $1/r^{D-2}$  (let us call it a ``stringy'' behavior). On the other hand at large distance the gravitational backreaction goes like $1/r^{D-3}$ (point-like behavior).

Let us consider the open string in $AdS_5$ studied in section \ref{backreaction}. 
In the region $z \gg r \gg z_0$ the geodesic distance away from the string is small and the backreaction goes like $1/r$: the backreacting string behaves like an extended object in five-dimensional flat spacetime.
To compute the gravitational potential energy between two open strings, we put a probe string in the gravitational field created by a backreacting string. 
The important observation is that the probe string always intersect the region of spacetime defined by $z \gg r \gg z_0$, where the backreacting string behaves like an extended object (see figure \ref{2strings}). 
When the euclidean distance $r$ separating the open strings is large, the contribution to the gravitational potential energy from this region dominates.
This explains why the gravitational potential energy between open strings always has a ``stringy'' behavior, even at large distance. It also implies that a point-particle approximation cannot reproduce this potential energy. 
Similar remarks apply to the analyses of sections \ref{generalizationDim} and \ref{generalization}.

Next we discuss the validity of the probe-brane approximation.
To this end we compare the formulas \eqref{resultnc}, \eqref{resultc} with the results one would obtain by treating the D3-branes in a probe approximation.
In that case we can safely approximate the open strings by point-particles.
Then the computation should be performed using the D3-branes DBI action coupled to the bulk gravity.

For probe D3-branes in D-dimensional flat space, the gravitational potential energy between two open strings of mass $m_0$ would be: 
\be E_{probe}(r) = -G_N^{(D)} \frac{m_0^2}{r^{D-3}} \ee
This is qualitatively very different from the result \eqref{resultnc}.
The decay of the gravitational potential energy at large $r$ is faster, and the equivalence principle is valid.
We deduce that in that case, the probe-brane approximation is never valid to discuss gravitational interactions between open strings.

If we compactify the space transverse to the D3-branes,
the probe-brane analysis leads to the following gravitational potential energy: 
\be E_{probe}(r) =  -\frac{G_N^{(D)}}{V_c}   \frac{m_0^2}{r}\ee
Comparing with our result \eqref{resultc}, we deduce that the probe brane approximation is justified only if the 't Hooft coupling is small enough.
This conclusion requires an extrapolation of our results at weak 't Hooft coupling.
The D3-branes' backreaction induces a correction to the Newtonian potential energy that violates the equivalence principle.


\subsection{Stringy violation of the equivalence principle and phenomenology}

As we mentioned several times already, the equivalence principle is violated in the four-dimensional theory living on D3-branes.
Indeed from equations \eqref{resultnc} and \eqref{resultc} it appears that the gravitational potential energy between two parallel open strings is not proportional to the inertial masses of the strings.

This effect follows from the fact that the open strings have a large extension in the bulk. As is clear from Figure \ref{2strings} the gravitational field varies over distances that are smaller than the open string's length. This is the reason why, even if the bulk theory obeys the equivalence principle, the lower-dimensional theory living on the D3-branes does not. 

Violations of the equivalence principle in brane-world scenarios have been previously noticed in
\cite{Dubovsky:2001pe}\cite{Kiritsis:2001bc}. In these papers the violation of the equivalence principle came from the finite thickness of the brane-world under consideration.
The picture we uncovered in the present work is not unrelated since we can identify the brane's thickness with the extension of the open strings in the bulk.

We have shown that the D3-branes' backreaction induces a large-distance interaction localized near the endpoints of parallel open strings, leading to a violation of the equivalence principle.
This effect is specific to string theory. It would not be captured by a point-particle approximation of the open strings.
It is a macroscopic manifestation of the microscopic differences between strings and point particles.
As such, it may lead to a way to confront a set of phenomenological vacua of string theory with experiment. 

Admittedly there is a large gap between the simple set-ups we considered in this paper, and D-branes constructions of phenomenological interest.
To fill this gap it would be useful to generalize our computations to higher-dimensional Dp-branes and intersections thereof that generically appear in model building.
It would also be interesting to study set-ups which provide confining gauge theories.


\subsection{Bulk interactions localized on branes}

In this work we showed that the gravitational potential energy between parallel open strings attached to D3-branes goes like $1/r$.
This is the expected behavior for a massless interaction in four-dimensional flat space.
However our results hold even if spacetime has more than four non-compact dimensions.

Localizing bulk interactions (and especially gravity) on branes is a problem that received a lot of attention.
One possibility is to rely on the existence of a normalizable collective mode of the graviton, as in the Randall-Sundrum scenario \cite{Randall:1999vf} (see e.g. \cite{Rubakov:2001kp} for a review). However it is difficult to obtain such a mode in a non-compact space \cite{Kiritsis:2006ua}\cite{Nitti:2008mn}.
Another way is to generate a four-dimensional metastable graviton as a resonance of the bulk gravitational field. 
This can be achieved if the action that describes spacetime effective field theory contains a localized four-dimensional Einstein-Hilbert term \cite{Dvali:2000hr}. In string theory such terms are known to appear, either as $\alpha'$-corrections to the DBI action \cite{Corley:2001hg} (however for BPS branes these terms vanish \cite{Bachas:1999um}), or as dimensional reduction of ten-dimensional $R^4$ couplings \cite{Antoniadis:2002tr}.
Generically some severe tuning is required for the lifetime of the metastable graviton to be long enough.

The mechanism at work in our case is quite different from the Randall-Sundrum or DGP-like scenarios.
The open strings essentially behave like infinitely extended objects from the point of view of the bulk interactions.
This results in a slower decay of the gravitational attraction at large distances.

An important open question is to derive an effective four-dimensional field theory description of this interaction.
At first sight such a description would differ significantly from Einstein's gravity coupled to the D3-branes gauge theory,
 since the potential energy we derived is not proportional to the inertial masses of the open strings.


\subsection{From general relativity to string theory}

In our computation we used a field theory description of gravity in the bulk. Thus we may ask about the corrections our results receive in string theory.

First let us mention that the mass \eqref{massString} of the open strings we are considering can be arbitrarily small. 
The computation of the gravitational potential energy in section \ref{potential} is valid when the distance $r$ separating the strings is greater than the length $z_0$ which is related to the mass of the string $m_0$.
This constraints reads:
\be r \gg \frac{R^2}{2\pi \alpha' m_0} \ee
For a ``microscopic'' string with a mass of order $\epsilon /\sqrt{\alpha'}$, with $\epsilon$ a small number, this implies that $r$ has to be much greater than the product between the string length and the 't Hooft coupling $\sqrt{\alpha'}g_s N$, divided by $\epsilon$. Since we assume that $r$ is a macroscopic distance this constraint is satisfied.

Next let us consider the issue of $\alpha'$ corrections.
We started with an anti-de Sitter spacetime which is weakly curved with respect to the string scale $\alpha'$. 
After backreaction of an open string, spacetime is still weakly curved in the region where the linear treatment of the backreaction is reliable. 
In the computation of the gravitational potential energy we only used the linear backreaction of the open string.
This gives confidence in the field theory approximation to bulk gravity.
%
Nevertheless a worldsheet derivation of the results we obtained would be most useful, as it would give a better control over $\alpha'$ corrections.
An important question could then be addressed, namely what is the fate of the gravitational interaction at weak 't Hooft coupling.

We do not have the ambition to initiate a worldsheet analysis in the present work, but
let us mention one reason why the string theory analysis is rather involved.
The four-dimensional gravitational effect we identified is tightly linked to the backreaction of the D3-branes.
From the worldsheet point of view the D-branes' backreaction takes the form of local counterterms that have to be added to the worldsheet conformal field theory in order to cancel open string tadpoles \cite{Fischler:1986ci}\cite{Fischler:1986tb}\cite{Keller:2007nd}. These counterterms trigger a renormalization group flow, which IR fixed point is the worldsheet theory after backreaction of the D-branes. 
Constructing this conformal field theory is an interesting but non-trivial problem (cf e.g. \cite{Berenstein:1999jq}).


\subsection{Reminiscence of the holographic principle}

Finally we would like to conclude with some speculative comments about holography.
We found that the gravitational interactions between open strings on D3-branes have a rather surprising manifestation.
The gravitational potential energy between parallel open strings is not proportional of the inertial masses of the strings.
For an observer living on the D3-branes, the resulting interaction is presumably quite far from what could be interpreted as gravity.
Moreover the behavior in $1/r$ obtained for the potential energy in \eqref{resultnc} is typical of a four-dimensional massless interaction.

This effective dimensional reduction on horizons is reminiscent of the holographic principle introduced in \cite{'tHooft:1993gx}\cite{Susskind:1994vu}.
The AdS/CFT correspondence \cite{Maldacena:1997re}\cite{Aharony:1999ti} provided a concrete realization of the holographic principle in string theory.
The current work is placed outside of the usual AdS/CFT framework since the open string theory is coupled to bulk gravity.
The fact that we nevertheless obtain a picture that seems compatible with the holographic principle is encouraging.
It stresses the importance of finding an effective field theory description of the gravitational interactions between open strings.


\section*{Acknowledgments}

This work is dedicated to A. and A. on the occasion of their first birthday.

The author would like to thank Costas Bachas, Cyril Closset, Aldo Cotrone, Emiliano Imeroni, Oleg Lunin, Francesco Nitti, Giuseppe Policastro,  Jan Troost, Pedro Vieira, Riad Ziour, and especially Ben Craps, Rob Myers, Amitabh Virmani and an anonymous referee for useful discussions and correspondence at various stages of this project.

This work is supported in part by the Belgian Federal Science Policy Office through the Interuniversity Attraction Pole P6/11, and in part by the "FWO-Vlaanderen" through the project G.0114.10N.


\begin{appendix}

\section{A bit of intuition from Newtonian gravity}\label{Newton}
In this appendix we briefly present elementary and well-known facts that may shed some light on the results contained in the bulk of this paper.

\subsection*{In five dimensions}

\begin{figure}
\centering
\includegraphics[scale=1]{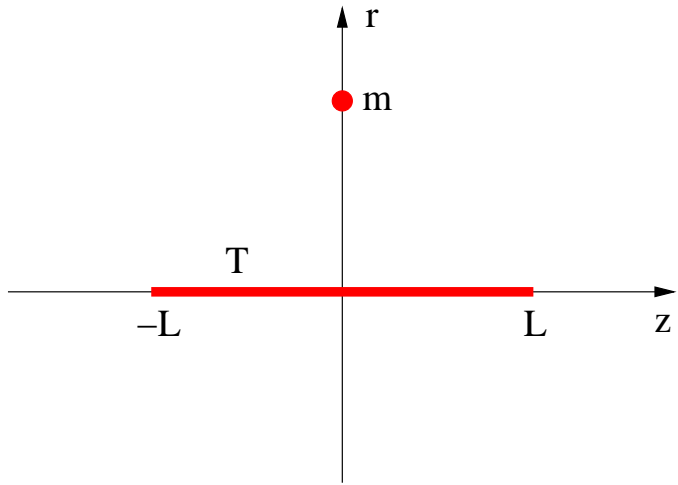}
\caption{A massive string with constant tension $T$ extends along the $z$ direction in a $D$-dimensional flat space. In appendix \ref{newton} we estimate the gravitational potential energy between this string and the point-mass $m$.
\label{newton}}
\end{figure}

We consider a static string in  five-dimensional flat spacetime. We assume that the string extends along the $z$ coordinate (cf Figure \ref{newton}). The string has length $2L$ and constant tension $T$. We denote by $r$ the transverse distance to the string. 
We place a point mass $m$ on the constant-$z$ hyperplane that crosses the string in its middle. The Newtonian potential energy between these two objects is easily evaluated as:
\be\label{newton5D} E(r) = -\int_{-L}^L dz G_N^{(5)} T m \frac{1}{r^2 + z^2} = -2 G_N^{(5)} T  m\frac{ \arctan\left( \frac{L}{r} \right)}{r} \ee
where $G_N^{(5)}$ is the five-dimensional Newton's constant.

At larges distances ($r \gg L$) the string behaves like a point-like object of mass $T L$:
\be E(r) \approx  -2 G_N^{(5)} T L m \frac{1}{r^2} \ee
On the other hand close to the string, namely for $r \ll L$, the potential has a four-dimensional behavior:
\be\label{potClose5D} E(r) \approx  -\pi G_N^{(5)} T m \frac{1}{r} \ee
In that case it does not depend on the mass of the string but only on its tension.
If we send the string length to infinity, equation \eqref{potClose5D} is valid at arbitrary large distances away from the string.

When a string touches a horizon its proper length typically goes to infinity. Thus from the previous discussion we can expect that the decay of the gravitational field created by this string at large distances will be slower than in flat space.
The computation of section \ref{backreaction} renders this intuition precise.

In order to compare with the results obtained in the bulk of the paper for open strings attached to D-branes (equation \eqref{Enc}), let us write down explicitly the gravitational potential energy between two parallel strings of size $2L$ and of constant tension in five-dimensional flat spacetime:
\beq 
r \gg L &:& E(r) \approx -2 G_N^{(5)} T^2 L^2 \frac{1}{r^2} \cr
r \ll L &:& E(r) \approx -\pi G_N^{(5)} T^2 L \frac{1}{r}
\eeq

\subsection*{Generalization to D dimensions}

Let us consider the same set-up in a D-dimensional flat spacetime, with $D\ge 5$. The Newtonian potential between this string and the point mass is, up to numerical factors:
\be r \gg L \ :\qquad E(r) \approx  - G_N^{(D)} T L m \frac{1}{r^{D-3}} \ee
\be r \ll L \ :\qquad E(r) \approx  - G_N^{(D)} T m \frac{1}{r^{D-4}} \ee

\end{appendix}

\end{document}